# A Novel And Efficient Bilateral Remote User Authentication Scheme Using Smart Cards

Al-Sakib Khan Pathan, *Student Member, IEEE,* Choong Seon Hong, *Member, IEEE,* and Tatsuya Suda, *Member, IEEE*

*Abstract*--This paper proposes a novel remote user authentication scheme using smart cards which allows both the authentication server (AS) and the user to verify each other's authenticity. Our scheme is efficient enough to resist the known attacks that could be launched against remote user authentication process.

## I. INTRODUCTION

With the proliferation of distributed computing, remote user authentication has become an important task in many applications like e-commerce, e-banking etc. As the communications in the authentication process are considered to occur over insecure channels, there is the chance of leaking out secret information and thus causing serious harm to the user. In this paper, we propose a novel and efficient remote user authentication scheme using smart cards which allows both the AS and the user to verify each other.

The rest of the paper is organized as follows: Section II states the related works, Section III mentions the preliminaries, Section IV presents our scheme, Section V contains a brief analysis, and Section VI concludes the paper.

## II. RELATED WORKS

[1] proposed a remote user authentication scheme based on the ElGamal's public key cryptosystem. [2] increased the efficiency of [1] by reducing the computation and communication costs. [3] showed that, [1] is vulnerable to masquerading attack. Later, in [4] the authors showed a different type of attack on [1] and proposed a modified and enhanced scheme for making it resistant against the known attacks. In 2003, [5] showed that, [2] is vulnerable to offline and online password guessing attacks and [6] is vulnerable to parallel session attack.

Most of the works mentioned above ensure that the AS could verify the authenticity of the user but the user cannot verify the validity of the AS. Hence, our scheme aims at achieving bilateral verification maintaining robust security of the scheme so that it could be resistant to the known attacks in remote user authentication process using smart cards.

This work was supported by MIC and ITRC projects. Dr. C.S. Hong is the corresponding author.

Al-Sakib Khan Pathan is a graduate student and research assistant at the Department of Computer Engineering, Kyung Hee University, South Korea. (e-mail: spathan@networking.khu.ac.kr, phone: +82 31 201-2987)

Dr. Choong Seon Hong is a professor at the Department of Computer Engineering, Kyung Hee University, South Korea. (e-mail: cshong@khu.ac.kr, phone: +82 31 201-2987)

Dr. Tatsuya Suda is a professor at the School of Information and Computer Science, University of California, Irvine, USA. (e-mail: suda@ics.uci.edu, phone: 949-824-5474)

## III. PRELIMINARIES

We use LU decomposition [7] for our scheme. LU decomposition is a procedure for decomposing a square matrix A ($N \times N$) into a product of a lower triangular matrix L and an upper triangular matrix U, such that,

$$A = LU \qquad (1)$$

Where, lower triangular matrix L and upper triangular matrix U have the forms,

$$L_{ij} = \begin{cases} l_{ij} & for \ i \geq j \\ 0 & for \ i < j \end{cases} \qquad U_{ij} = \begin{cases} u_{ij} & for \ i \leq j \\ 0 & for \ i > j \end{cases}$$

So, for a square matrix of $4 \times 4$, equation (1) looks like:

$$\begin{bmatrix} l_{11} & 0 & 0 & 0 \\ l_{21} & l_{22} & 0 & 0 \\ l_{31} & l_{32} & l_{33} & 0 \\ l_{41} & l_{42} & l_{43} & l_{44} \end{bmatrix} \cdot \begin{bmatrix} u_{11} & u_{12} & u_{13} & u_{14} \\ 0 & u_{22} & u_{23} & u_{24} \\ 0 & 0 & u_{33} & u_{34} \\ 0 & 0 & 0 & u_{44} \end{bmatrix} = \begin{bmatrix} a_{11} & a_{12} & a_{13} & a_{14} \\ a_{21} & a_{22} & a_{23} & a_{24} \\ a_{31} & a_{32} & a_{33} & a_{34} \\ a_{41} & a_{42} & a_{43} & a_{44} \end{bmatrix} \qquad (2)$$

According to the definition, elementary matrix $E$ is an $N \times N$ matrix if it can be obtained from the identity matrix $I_n$ by using one and only one elementary row operation (e.g., elimination, scaling, or interchange) [8], [10]. Elementary row operations are, $R_i \leftrightarrow R_j$, $cR_i \leftrightarrow R_i$, $R_i + cR_j \leftrightarrow R_i$. If the elementary matrices corresponding to the row operations that we use are, $E_1, E_2 \ldots E_k$, then, $E_k \ldots E_2 E_1 A = U$. Hence, $A = (E_k \ldots E_2 E_1)^{-1} U$ or $L = E_k^{-1} \ldots E_2^{-1} E_1^{-1}$

## IV. OUR BILATERAL AUTHENTICATION SCHEME

### A. Pre-processing - Symmetric Matrix Creation

At first, a secret symmetric key matrix A (dimension $N \times N$) is generated by the AS, where $N$ is the maximum number of users that could be supported. Each element $A_{ij}$ of A is assigned a distinct key from a key pool (generated earlier) such that, $A_{ij} = A_{ji}$ for, i, j = $\overline{1, N}$. Then, LU-decomposition is applied on the matrix A to get L and U.

### B. Details of Our Scheme

Our scheme has mainly two phases – User Registration Phase, Login & Bilateral Authentication Phase.

**User Registration Phase:** We assume that, this phase occurs over a secure channel. Let $F_h$ be a secure one-way hash function [9]. In the registration phase, the user $U_a$ with identity $ID_a$ first submits his identity ($ID_a$) and arbitrarily chosen password $PW_a$ to the AS for registration. In turn the AS does the following steps:

1. Generates two random numbers $x$ and $y$ within the range N (the dimension of the matrices).

2. It selects the $x$th row from L matrix $L_R(x)$, $x$th column from U matrix $U_C(x)$, and $y$th column from U matrix $U_C(y)$.

3. Computes $L_R(x) \times U_C(y) = K_{xy}$ and $\theta = F_h(ID_a \oplus K_{xy}) \oplus PW_a$.

4. Issues a smart card containing $(F_h, K_{xy}, \nu, U_C(x), \theta)$, to the user, where $\nu = (\varphi \oplus y)$ with $\varphi$ is an arbitrary number which is kept secret and owned by the authentication server.

**Login & Bilateral Authentication Phase:** When the user needs to login, he attaches the smart card to the input device and keys in his identity $ID_a$ and password $PW_a$. The smart card performs the following operations:

1. Generates a random number $r$ with the same length of $K_{xy}$ and computes $H_a = K_{xy} \oplus F_h(r)$, and $S_a = \theta \oplus PW_a \oplus r$.

2. Sends the login request message, $M = (ID_a, H_a, \nu, U_C(x), S_a, T)$, (here, T is the current timestamp) to the AS.

After getting M, the AS performs the operations:

1. Checks the validity of $ID_a$. If the format is different than the allowed format, it rejects the request.

2. Tests the time interval $(T' - T) \leq \Delta T$, where $T'$ is the timestamp of receiving the message M and $\Delta T$ is the maximum allowed time interval for transmission delay. If $\Delta T$ is greater than its boundary condition, the request is rejected.

3. Now AS computes, $(\nu \oplus \varphi)$ which eventually generates the value of $y$. AS now knows which row is to be selected from the L matrix for this user and selects the $y$th row $L_R(y)$ and computes, $L_R(y) \times U_C(x) = K_{yx}$

4. Computes $t = F_h(ID_a \oplus K_{yx})$, and $r' = t \oplus S_a$.

5. Computes, $H_a \oplus F_h(r')$, which is expected to generate the value of $K_{xy}$ for a legitimate user, as $r'$ should be equal to the randomly generated number in the user side, $r$.

6. Now the server checks whether the condition, $K_{xy} = K_{yx}$ holds or not. If it does not hold, the server detects the user $U_a$ as an invalid user otherwise as a valid user. For the invalid users, the server rejects the login request and for the legitimate users, it proceeds to the next steps.

7. Computes $M' = F_h(K_{yx}\ X\text{-}NOR\ T'')$, where $T''$ is the current timestamp, *X-NOR* means the bitwise Exclusive-NOR (XNOR) operation and sends $(M', T'')$ to the user $U_a$.

Upon receiving this message from the AS, the user $U_a$,

1. Verifies the boundary condition, $T'''-T'' \leq \Delta T$, where $T'''$ is the timestamp of receiving the message.

2. Then it computes, $F_h(K_{xy}\ X\text{-}NOR\ T'')$ and if it equals to the received $M'$, the user verifies the legitimacy of the AS.

As a symmetric matrix is used for LU-decomposition, $K_{xy}=K_{yx}$ and the procedure works for the legitimate authentication server and user. Thus, we ensure bilateral verification in our scheme.

## V. SECURITY AND PERFORMANCE ANALYSIS

In this section we analyze our scheme in brief.

**Replay Attacks:** Replaying an old login request, $(ID_a, H_a, \nu, U_C(x), S_a, T)$ could not do any harm as, in the second step of the login & authentication phase in the AS, this will be rejected.

**Masquerade Attack:** Our scheme is resistant to masquerading attack presented in [3] as, our algorithm does not derive any password based on the identity of the user, rather it is chosen by the user and an attacker must have pre-stored other information to masquerade a legitimate user.

**Eavesdropping:** By eavesdropping, an attacker could get little information that could be useful. In fact, even if the attacker knows $U_c(x)$, without knowing the corresponding row information or other information, it could not do any harm.

**Forging Attack:** No attacker could forge a valid login request as it will be detected by the AS during the authentication process.

**Other Attacks:** Offline and online password guessing attacks and parallel session attacks mentioned in [5] could not be performed against our scheme as the bilateral authentication depends on some pre-stored information.

As in each step different message formats are used, the attacker cannot take advantage of similar previous message to use it in any later step. No attacker can compute the hash outputs as $F_h$ is a one-way hash function and it is kept secret. $\varphi$ is used for hiding the required row number even from the user and provides effective security for our algorithm. There is no exponential operation in our scheme; hence, the computation is fairly easy. For supporting more users, multiple symmetric matrices could be used.

## VI. CONCLUSION

In this paper, we have proposed a novel remote user authentication scheme using smart cards which ensures bilateral authentication so that both the parties participating in the process could verify each other's validity.

REFERENCES

[1] Hwang, M.-S. and Li, L.-H., "A New Remote User Authentication Scheme Using Smart Cards", IEEE Transactions on Consumer Electronics, Vol. 46, No. 1, February, 2000, pp. 28-30.
[2] Sun, H.-M., "An Efficient Remote User Authentication Scheme Using Smart Cards", IEEE Transactions on Consumer Electronics, Vol. 46, No. 4, November, 2000, pp. 958-961.
[3] Chan, C.-K. and Cheng, L. M., "Cryptanalysis of a Remote User Authentication Scheme Using Smart Cards", IEEE Transactions on Consumer Electronics, Vol. 46, No. 4, November, 2000, pp. 992-993.
[4] Shen, J.-J., Lin, C.-W., Hwang, M.-S., "A Modified Remote User Authentication Scheme Using Smart Cards", IEEE Transactions on Consumer Electronics, Vol. 49, No. 2, May, 2003, pp. 414-416.
[5] Hsu, C.-L., "Security of Two Remote User Authentication Schemes Using Smart Cards", IEEE Transactions on Consumer Electronics, Vol. 49, No. 4, November, 2003, pp. 1196-1198.
[6] Chien, H. Y., Jan, J. K., and Tseng, Y. M., "An Efficient and Practical Solution to Remote Authentication: Smart Card", Computers and Security, Vol. 21, No. 4, 2002, pp. 372-375.
[7] Zarowski, C. J., "An Introduction to Numerical Analysis for Electrical and Computer Engineers", Hoboken, NJ John Wiley & Sons, Inc. (US), 2004, pp. 148-151.
[8] Nakos, G., and Joyner, D., Linear Algebra with Applications, Brooks/Cole USA, 1998, pp. 188-194.
[9] National Institute of Standards and Technology, NIST FIPS PUB 180, "Secure hash standard", U.S. Department of Commerce, 1993.
[10] Birkhauser, Linear Algebra, Birkhauser Boston, 1997, pp. 33-37.